\documentclass[preprint,10pt]{elsarticle}

\usepackage{amsmath,amssymb}
\usepackage{graphicx}
\usepackage{booktabs}
\usepackage{siunitx}
\usepackage{hyperref}
\usepackage{natbib}
\biboptions{sort&compress}
\usepackage{tikz}
\usepackage{tabularx}
\usepackage{array} % optional but useful
\usetikzlibrary{positioning,arrows.meta}
\usepackage{booktabs}
\usepackage{tabularx}
\usepackage{enumitem}
\usepackage{ragged2e} % for \RaggedRight in X columns
\newcolumntype{Y}{>{\RaggedRight\arraybackslash}X}

\begin{document}

% ---------------- Front matter ----------------
\begin{frontmatter}

\title{Horizon-Aware Forecasting of Passenger Assistance Demand for Rail Station Workforce Planning}

\author[inst1]{Dr. Michael Patrick Sheehan\corref{cor1}}
\ead{michael.sheehan@lner.co.uk}

\author[inst1]{Irina Timoshenko}
\ead{irina.timoshenko@lner.co.uk}

\address[inst1]{London North Eastern Railway, Digital and Innovation Department, York, UK}

\cortext[cor1]{Corresponding author}

\begin{abstract}
Passenger assistance services are essential for accessible rail travel, yet demand varies substantially across stations and over time, creating challenges for workforce planning and staff rostering. This paper presents a data-driven decision support framework for forecasting station-level passenger assistance demand and translating forecasts into workforce plans. The forecasting component applies a horizon-aware Prophet modelling approach using multi-source operational data, while the planning component maps demand forecasts to staffing requirements under service and operational constraints through an interpretable red--amber--green risk framework. The approach has been implemented within a production-grade system to support routine planning and staffing decisions across LNER-managed stations. Results demonstrate improved forecast accuracy relative to year-on-year baseline methods, with absolute error reduced by up to $76.9\%$, and show that forecast-informed staffing is associated with an approximate $50\%$ reduction in failed passenger assistance deliveries attributable to staff availability. These findings highlight the value of integrating interpretable forecasting with operational workforce planning to improve accessibility service reliability in rail operations.
\end{abstract}


\end{frontmatter}

% ---------------- Main text ----------------

\section{Introduction}
\label{sec:intro}

Passenger assistance services play a critical role in enabling accessible rail travel and ensuring impartial access to the rail network for passengers who require additional support. Across many rail systems, operators are subject to regulatory obligations to provide timely and reliable assistance, with performance monitored by oversight bodies such as the Office of Rail and Road (ORR). Failure to deliver booked or expected assistance can have significant consequences, including degraded passenger experience, reputational damage, and direct financial costs arising from compensation, service recovery, and reactive staffing interventions. Recent national reporting indicates that only around $78\%$ of passengers receive all of the assistance they booked \cite{orr_passenger_assist_2025}, highlighting persistent delivery challenges. As demand for passenger assistance continues to grow, driven by demographic change and increased awareness of assistance services, the ability to anticipate demand accurately has become an increasingly important operational challenge.

From an operational perspective, passenger assistance demand exhibits substantial temporal and spatial variability across stations.
Volumes are shaped by strong weekly and seasonal patterns, but are also influenced by external factors such as weather conditions, service disruption, special events, and evolving booking behaviours.
Despite this complexity, assistance demand forecasting in practice often remains highly manual, relying on historical averages or simple year-on-year growth assumptions applied to previous periods.
Such approaches are labour intensive, slow to update, and insufficiently responsive to short-term changes in demand or emerging booking patterns, limiting their effectiveness for proactive workforce planning.

The limitations of manual and static forecasting approaches are particularly acute when staffing decisions must be made under uncertainty.
Underestimation of assistance demand can lead to inadequate frontline staffing levels, increasing the risk of service failure, while overestimation can result in inefficient deployment of staff and increased operational costs.
Effective workforce planning therefore requires not only accurate demand forecasts, but also a systematic mechanism for translating forecasts into staffing insights that support operational decision-making.

This paper addresses these challenges by presenting an integrated, data-driven decision support system for predicting passenger assistance demand and supporting operational workforce planning at the station level. The proposed framework leverages the strong temporal structure present in assistance demand data while incorporating external regressors that capture booking activity and environmental conditions. Forecast outputs are then operationalised through a workforce planning tool that translates predicted demand into staffing requirements and provides a simple risk-based assessment of staffing adequacy using a red--amber--green (RAG) classification. The framework is designed explicitly around operational planning horizons and has been deployed within routine planning workflows to support staffing decisions in live use.

The contributions of this paper are fourfold:
\begin{itemize}
    \item We develop a production deployed support system for passenger assistance workforce planning that integrates station-level demand forecasting with an operational staffing assessment layer.
    \item We introduce a horizon-aware forecasting framework that accounts for the changing availability of booking and operational information across planning lead times, aligning model structure with the decision horizons at which workforce actions are taken.
    \item We formulate an interpretable method for translating demand forecasts into staffing insights by linking predicted demand to effective hourly workforce capacity and expressing staffing adequacy through a Red--Amber--Green classification.
    \item We demonstrate measurable operational benefits in live use, including substantial improvements in forecast accuracy relative to year-on-year planning methods and an associated reduction in failed passenger assistance deliveries attributable to staff availability.
\end{itemize}

The paper positions passenger assistance forecasting not simply as a statistical prediction problem, but as an operational decision-support challenge in which the value of forecasting lies in its ability to improve staffing actions and accessibility service reliability. The remainder of this paper is structured as follows. Section~\ref{sec:related} reviews relevant literature on passenger demand forecasting and workforce planning in transport operations. Sections~3 and~\ref{sec:planning} present the forecasting methodology and workforce planning approach, respectively. Results are presented in Section~\ref{sec:results}, followed by a discussion in Section~\ref{sec:discussion} and conclusions in Section~\ref{sec:conclusion}.

\section{Background and related work}
\label{sec:related}

\subsection{Passenger assistance in rail operations}
Passenger assistance services support passengers who require additional help to complete their rail journeys, including assistance with boarding and alighting, navigation within stations, and interchanging between services.
In many rail systems, assistance can be requested in advance through a booking process or provided on the day of travel through turn-up-and-go (TUAG) arrangements.
Pre-booked assistance allows operators to plan staff deployment ahead of time, while TUAG services require sufficient staffing capacity to respond to demand that is not known in advance.

Passenger assistance is delivered primarily at the station level by frontline staff.
Responsibility for managing and delivering assistance services typically sits with the train operating company (TOC) that manages or co-manages a station.
For example, along the East Coast Main Line, London North Eastern Railway (LNER) manages or co-manages thirteen stations, each with distinct operational characteristics and passenger profiles.
Within stations, assistance delivery may be handled by staff whose primary role is passenger assistance, as well as by staff whose primary responsibilities lie elsewhere (e.g.\ dispatch or platform duties) but who can provide assistance when operational conditions allow.
This layered responsibility structure introduces additional complexity for workforce planning, as staff availability for assistance is contingent on wider station operations.

\subsection{Characteristics of passenger assistance demand}
\begin{figure}[!htbp]
    \centering
    \includegraphics[width=0.9\linewidth]{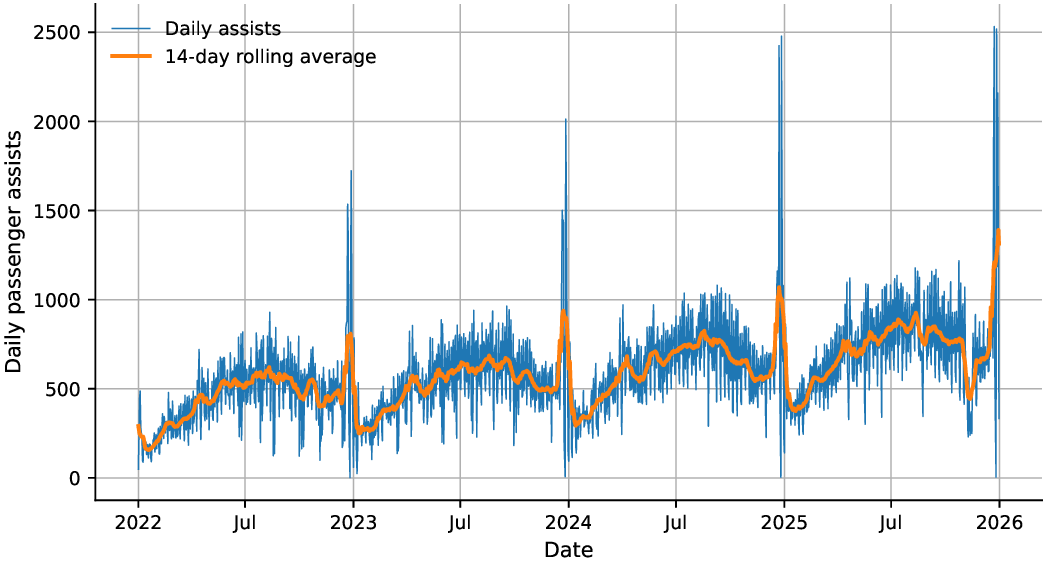}
    \caption{Station-level passenger assistance requests over the last four years, illustrating strong seasonality and peak demand periods.}
    \label{fig:assist_timeseries}
\end{figure}

Passenger assistance demand exhibits pronounced temporal and spatial variation across stations.
Strong day-of-week and seasonal patterns are observed, reflecting broader travel behaviour, school holidays, and public holiday periods.
Demand is also highly heterogeneous across stations, with major interchange and long-distance stations experiencing substantially higher and more variable volumes than smaller or commuter-focused stations. In addition to regular seasonal effects, assistance demand can be volatile and subject to sharp spikes.
Periods such as the Christmas and end-of-year holiday season are characterised by particularly high volumes of assistance requests, creating sustained peaks that place pressure on station staffing.
Figure~\ref{fig:assist_timeseries} illustrates passenger assistance request volumes over the last four years, highlighting recurring seasonal patterns, exceptional peak periods and year-on-year trend growth.

\begin{figure}[htbp!]
    \centering
    \includegraphics[width=0.9\linewidth]{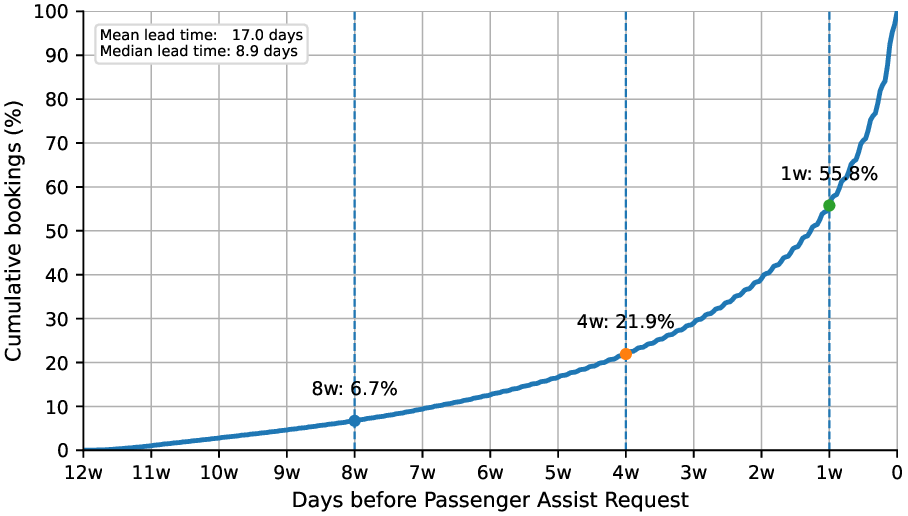}
    \caption{Cumulative distribution of passenger assistance booking lead times.}
    \label{fig:booking_leadtime}
\end{figure}

While assistance bookings can typically be made up to twelve weeks in advance, booking lead-time behaviour indicates that a substantial proportion of requests are made close to the date of travel.
Figure~\ref{fig:booking_leadtime} shows the cumulative distribution of booking lead times, demonstrating that approximately $44\%$ of bookings occur within the final week before service.
This concentration of late bookings limits the time available for workforce adjustments, particularly when staff rosters have already been finalised.

\subsection{Workforce planning and operational constraints}
Workforce planning for passenger assistance operates across multiple overlapping time horizons, each associated with different decision-makers and degrees of flexibility.
At longer horizons, base rosters are typically set several weeks in advance, with roster clerks responsible for allocating staff to shifts approximately four weeks ahead of operation.
At intermediate horizons, accessibility or assistance planning teams review anticipated demand and adjust plans around two weeks before travel, for example through targeted redeployment or overtime planning.
At shorter horizons, for instance a few days before departure, service delivery and station managers are responsible for responding to emerging demand pressures through reactive measures such as rest day working at additional costs to the business.

These staggered planning horizons mean that forecast accuracy and timeliness have different operational value depending on when information becomes available.
Late surges in bookings may fall outside the window in which roster adjustments are feasible, increasing reliance on reactive management and potentially affecting service reliability.
Consequently, workforce planning for passenger assistance must balance early forecasts with the ability to update expectations as new information becomes available.

\subsection{Current forecasting and planning practices}
\label{sec: workforce planning and constraints}
In practice, passenger assistance workforce planning often relies on a combination of base staffing assumptions and simple demand estimates.
Base rosters are typically designed to cover expected average demand, with additional staffing considered during known peak periods such as major holiday seasons, when assistance volumes are historically high.
Outside these periods, forecasting practices are frequently informal and manual, involving the inspection of recent booking trends or comparison against previous years.

While such approaches provide a degree of operational awareness, they are inherently reactive and limited in their ability to account for evolving booking patterns, external influences, or short-term demand surges.
As a result, planners may have limited advance visibility of staffing risks, particularly when demand deviates from historical norms or when late booking behaviour intensifies.
This operational context motivates the need for systematic, data-driven forecasting approaches that can support workforce planning across multiple time horizons.

\begin{table}[!htbp]
\centering
\caption{Passenger assistance workforce planning horizons, responsibilities, and typical duties.}
\label{tab:planning_horizons}
\renewcommand{\arraystretch}{1.15} % slightly more vertical breathing room

\begin{tabularx}{\linewidth}{p{3cm} p{4cm} X}
\toprule
\textbf{Planning horizon} & \textbf{Responsible role/team} & \textbf{Typical duties} \\
\midrule

8--12 weeks ahead &
Strategic / resource planning &
Coordinate timetable and works impacts; set high-level station allocations; plan recruitment to close gaps. \\
\addlinespace

2--4 weeks ahead &
Roster clerks / accessibility planning teams &
Build rosters aligned to expected demand; manage leave, training, and planned overtime. \\
\addlinespace

Less than 1 week ahead &
Service delivery \& Station managers &
Deploy staff day to day; approve short-notice overtime or rest day working \\

\bottomrule
\end{tabularx}
\end{table}

\subsection{Related Work}

Research on passenger assistance and accessibility in rail has primarily focused on
service quality, regulatory oversight, and passenger experience rather than predictive
operational planning. In the UK, the Office of Rail and Road (ORR) regularly monitors assisted travel performance, highlighting persistent challenges in maintaining consistent satisfaction and reliability levels despite policy interventions \cite{orr_assisted_travel}. Such findings underline the operational importance of accurately anticipating assistance
demand to ensure compliance with accessibility standards and to support equitable access to the rail network. However, the literature contains comparatively limited work on quantitative forecasting of assistance volumes, with most studies emphasising service delivery evaluation rather than predictive workforce planning.
A broader body of research exists on passenger demand forecasting across the rail and wider transport industries. Traditional approaches in rail planning have relied on econometric models and elasticity-based frameworks, as reflected in industry guidance such as the Passenger Demand Forecasting Handbook \cite{pdfh}. More recent academic work has expanded this perspective through the use of machine learning and data-driven techniques to capture nonlinear relationships and short-term variability in demand \cite{passengerdemandscheduled,rail_ml_forecasting, clustering_demand}. These studies demonstrate that transport demand exhibits
strong temporal structures and is influenced by a combination of calendar effects, behavioural patterns, and exogenous variables such as weather and disruptions. Similar methodological trends are evident in aviation demand forecasting, where comparative reviews highlight the shift from classical statistical models towards hybrid and machine learning approaches capable of handling complex seasonal dynamics and
external regressors \cite{aviation_review1, aviation_review2}.

Within time-series forecasting, the Prophet model has gained significant attention as a
practical and interpretable framework for modelling data with multiple seasonalities and
calendar effects. Originally introduced as a decomposable additive model
\cite{prophet_original}, Prophet has been widely adopted in both academic and applied
contexts due to its ability to capture trend changes, holiday effects, and recurring
seasonal patterns while remaining robust to missing data and structural shifts.
Subsequent studies have demonstrated its effectiveness across a range of domains,
including customer behaviour modelling and short-term demand prediction
\cite{prophet_customer, prophet_methods}.

In the transport sector specifically, Prophet has been applied to passenger flow
prediction in urban transit, seaports, and multimodal travel systems, where strong weekly
and annual seasonalities are prevalent \cite{prophet_transport1, prophet_transport2,
prophetfreight}. These applications consistently report competitive performance
relative to alternative statistical and machine learning models, particularly when
interpretability and ease of deployment are important operational considerations. The
model’s capacity to incorporate external regressors and domain-specific holiday effects
makes it especially well suited to transport contexts characterised by pronounced
calendar-driven variability.

Despite this growing body of work, the application of decomposable time-series models to passenger assistance demand remains largely unexplored. While passenger demand forecasting has been studied extensively across multiple horizons, these approaches cannot be directly translated to passenger assistance, as assistance volumes are not simply a fixed proportion of passenger flows and vary substantially across stations depending on passenger demographics, journey characteristics, and local operating practices. Consequently, methods designed to predict aggregate demand may fail to capture the distinct temporal patterns and heterogeneity observed in assistance services. Existing studies therefore focus primarily on network-level passenger volumes rather than service-specific operational tasks such as assistance delivery. This paper contributes to the literature by bridging accessibility operations and demand forecasting, demonstrating how a Prophet-based framework can be adapted to model assistance demand at the station level and integrated directly into workforce planning processes. In doing so, it extends the use of interpretable time-series methods beyond traditional passenger volume forecasting toward operational decision support in accessible rail services.

\section{Problem definition and data}
\label{sec:data}

\subsection{Operational forecasting task}
We consider the problem of forecasting station-level passenger assistance demand to support workforce planning decisions.
The forecasting target is the hourly count of \textit{pre-booked} passenger assistance events at each station.
A passenger assistance request may involve multiple stations (e.g.\ an origin and destination, and potentially intermediate interchange stations).
In the source data, a journey-level request (e.g.\ KGX--EDB) is decomposed into the individual assistance events required at each relevant station, specifically a departure assist (\texttt{DEP}) at the origin station and an arrival assist (\texttt{ARR}) at the destination station; where a journey includes multiple legs, the decomposition extends to the corresponding \texttt{DEP}/\texttt{ARR} events for each leg.
This event-level representation aligns with the operational reality that assistance delivery and staffing requirements are station-based.

\subsection{Data sources}
Passenger assistance data are provided by the Rail Delivery Group (RDG) and cover requests across all stations and TOCs.
The dataset is supplied in anonymised form with personally identifiable information removed, and records are deduplicated prior to analysis.
Each record includes timestamps required to represent (i) when the booking was created and (ii) the scheduled service times associated with the assistance event (departure or arrival). In addition, we incorporate meteorological covariates by joining station-level weather observations to the passenger assistance time series.
Weather features include temperature, rainfall (mm), and humidity, and are aligned to the same hourly resolution as the forecasting target.

\subsection{Data preparation and feature design}
A dedicated preprocessing pipeline is used to construct a continuous station-hour dataset suitable for time-series modelling and operational inference.
Key steps are summarised below.

\paragraph{Station filtering and deduplication.}
Records are filtered to the station set in scope and duplicate rows are removed to mitigate repeated entries.
Events are then ordered chronologically to support time-based feature construction.

\paragraph{Filtering to pre-booked assistance only.}
We restrict the modelling dataset to pre-booked assistance events.
Turn-up-and-go (TUAG) records can be systematically distorted during days of disruption, where pre-booked assistance may be reassigned to alternative services and subsequently recorded as TUAG.
To avoid introducing disruption-driven labelling artefacts into the training target, only pre-booked events are retained for forecasting. Separately, a station-level seasonal TUAG rate, estimated from historical data, is applied to provide station delivery managers with an indicative estimate of expected TUAG assistance volumes under normal, non-disrupted operating conditions.

\paragraph{Hourly aggregation of assistance events.}
Passenger assistance events are aggregated to hourly counts for each station.
For hourly aggregation, the effective event timestamp is defined as the scheduled departure timestamp for \texttt{DEP} events and the scheduled arrival timestamp for \texttt{ARR} events.
This produces a target variable $y_{s,t}$ representing the number of assistance events at station $s$ during hour $t$.
A continuous station-hour index is constructed so that missing hours are explicitly represented with zero counts.

\paragraph{Booking lead-time features.}
To capture evolving booking behaviour, we construct a set of cumulative booking features that measure the volume of bookings created by specified lead-time thresholds.
For each station-hour $(s,t)$ and each threshold $\tau$ (e.g.\ 2 weeks before the scheduled event time), we compute the number of bookings with booking creation time less than or equal to $t-\tau$.
These \textit{as-of} features reflect what would have been known at different decision points and provide an operationally meaningful set of regressors for forecasting.
We additionally include difference features between adjacent thresholds (e.g.\ bookings between 2 and 1 weeks prior) to represent the pace of late-booking accumulation.

\paragraph{Weather covariates.}
Hourly weather observations are joined to the station-hour dataset.
Temperature, rainfall, and humidity are included as external regressors, aligned by station and hour.

The output of preprocessing is a station-hour panel with timestamps (\texttt{ds}), station identifiers, target hourly assistance counts (\texttt{y}), and a set of external regressors including booking lead-time features and weather covariates.

\paragraph{Feature engineering and data transformation.}
Prior to model fitting, numerical regressors are transformed to ensure comparability and stable estimation.
Continuous features, including booking lead-time variables and weather covariates, are scaled using a fitted transformation (standardisation by default), with the choice of scaling method applied consistently across all stations. For a small number of regressors with intermittent missing values, simple imputation is applied prior to scaling to preserve continuity in the time series. All transformation parameters are learned using the training data only and subsequently applied unchanged to the test and inference periods to avoid information leakage.
Model development follows a time-ordered training--test split, with the earliest 80\% of observations used for training and the remaining 20\% reserved for out-of-sample evaluation. To reflect real-world deployment conditions, models are trained and evaluated using only information that would have been available at the time the forecast is made. In particular, booking-derived regressors are constructed strictly on an \textit{as-of} basis relative to the forecast origin, ensuring that each observation contains only signals observable at the corresponding lead time. This prevents leakage of future information into the model and ensures that reported performance reflects realistic operational forecasting conditions rather than retrospective fitting.

\subsection{Horizon-bucket forecasting and routing}\label{sec:prophet}

Operational workforce planning requires forecasts at multiple lead times, ranging from several weeks ahead,supporting roster and resource planning, to short-horizon forecasts that inform near-term service delivery management.
A central challenge in this context is that the predictive value of booking-related signals varies substantially with forecast lead time as shown in Figure~\ref{fig:booking_leadtime}.
For dates far in the future, only a small proportion of eventual passenger assistance bookings have been created, whereas for near-term dates booking accumulation accelerates and provides a strong and rapidly changing signal.
To account for this horizon-dependent behaviour, we adopt a horizon-bucket forecasting approach in which separate models are estimated for different forecast lead-time ranges.

Let $t_0$ denote the forecast origin and $t$ a future timestamp, with the forecast horizon defined as $H = t - t_0$ (measured in days).
We define a set of non-overlapping horizon buckets $\mathcal{B}_k = [H_k^{\min}, H_k^{\max}]$ that collectively cover the operational planning window.
In this study, buckets correspond to very short-, short-,  medium I-, medium II- and longer-term horizons (e.g.\ $[1,2]$, $[3,7]$, $[8,14]$,$[15,28]$  and extended horizons beyond four weeks).
In practice, the framework can be easily adapted to suit more granular bucket definitions when finer horizon resolution is operationally or empirically beneficial.
For each bucket $\mathcal{B}_k$, a distinct Prophet model is trained using historical data, and future timestamps are deterministically routed to the appropriate bucket-specific model based solely on their horizon $H$.
Predictions from the bucket-specific models are subsequently concatenated to produce a single, continuous forecast trajectory.

Within each horizon bucket, passenger assistance demand is modeled using the additive decomposition
\begin{equation}
y(t) = g(t)
+ \sum\limits_{j \in \{\text{daily},\text{weekly},\text{yearly}\}} s_j(t)
+ h(t) + \mathbf{x}_k(t)^\top \boldsymbol{\beta}_k + \varepsilon(t),
\end{equation}
where $g(t)$ is a piecewise-linear trend component, $s_j(t)$ represent daily, weekly, and yearly seasonal effects, $h(t)$ captures holiday and special-period effects, $\mathbf{x}_k(t)$ is a vector of external regressors specific to horizon bucket $\mathcal{B}_k$ and $\varepsilon(t)$ is the remaining error term.
The inclusion of multiple seasonal components enables the model to capture intraday cycles, strong weekly structure, and longer-term seasonal variation observed in passenger assistance demand.

Holiday effects play a particularly important role in explaining extreme demand peaks, with the Christmas and year-end period being the most prominent example.
Although this period exhibits a strong recurring seasonal influence, associated passenger assistance demand is highly dynamic from year to year.
The timing, intensity, and duration of demand peaks vary depending on the day of the week on which Christmas falls, leading to shifts in travel patterns that may produce more concentrated peaks in some years and more dispersed demand in others.
As a result, demand surges may occur on different calendar days across years and cannot be adequately captured by fixed-date seasonal terms alone without inducing temporal misalignment.

To address this limitation, the model incorporates a set of domain-specific holiday windows around the Christmas and year-end period rather than treating Christmas as a single-day effect.
These windows include lead and lag intervals designed to capture pre-Christmas and post-Christmas travel behaviour, allowing the model to adapt to the timing and spread of demand peaks.
This approach enables the forecasting framework to represent both the strong recurring influence of the Christmas period and the year specific shifts in travel patterns driven by calendar effects.
Additional holiday windows are included for other high impact periods such as Easter and major public holidays.

The regressor set $\mathbf{x}_k(t)$ is tailored to each horizon bucket to reflect the information that would have been available to planners at the time forecasts are required.
For longer horizons, regressors primarily capture calendar structure and early booking indicators, while shorter horizons progressively incorporate richer booking lead-time features and short-term external signals such as weather conditions.
This horizon-aware specification allows the framework to exploit the increasing predictive power of near-term booking accumulation while maintaining stable and interpretable behaviour for longer-range forecasts. A summary of the horizon routing framework and the assoicated regressors for each horizon bucket are detailed in Table \ref{tab:horizon_buckets}.

\begin{table}[!htbp]
\centering
\caption{Horizon-bucket routing and horizon-specific regressor availability.}
\label{tab:horizon_buckets}
\small
\setlength{\tabcolsep}{5pt}
\renewcommand{\arraystretch}{1.12}

\begin{tabularx}{\linewidth}{p{2.1cm} c p{3.2cm} Y}
\toprule
\textbf{Horizon bucket} & \textbf{Days} & \textbf{Primary operational use} & \textbf{Additional regressors} \\
\midrule

\textbf{Very Short-term} & 1--2 &
Station Managers \& Deputy Team Leaders &
\begin{itemize}[leftmargin=*, nosep, topsep=0pt]
  \item Cumulative \textit{as-of} booking counts at fine lead-time thresholds ($\tau \leq 2$ days)
  \item Adjacent-threshold differences capturing late-booking accumulation
  \item Short-horizon meteorological covariates (forecast precipitation, temperature, humidity)
\end{itemize}
\\
\cmidrule(lr){1-4}

\textbf{Short-term} & 3--7 &
Near-term service delivery management &
\begin{itemize}[leftmargin=*, nosep, topsep=0pt]
  \item Cumulative \textit{as-of} booking counts at fine lead-time thresholds ($\tau \leq 1$ week)
  \item Adjacent-threshold differences capturing late-booking accumulation
  \item Short-horizon meteorological covariates (forecast precipitation, temperature, humidity)
\end{itemize}
\\
\cmidrule(lr){1-4}

\textbf{Medium-term I} & 8--14 &
Tactical workforce adjustments &
\begin{itemize}[leftmargin=*, nosep, topsep=0pt]
  \item \textit{As-of} booking counts at intermediate thresholds ($\tau \approx 2$--4 weeks)
  \item Adjacent-threshold differences
\end{itemize}
\\
\cmidrule(lr){1-4}

\textbf{Medium-term II} & 15--28 &
Short-range roster planning &
\begin{itemize}[leftmargin=*, nosep, topsep=0pt]
  \item \textit{As-of} booking counts at longer thresholds ($\tau \approx 4$--8 weeks)
  \item Adjacent-threshold differences
\end{itemize}
\\
\cmidrule(lr){1-4}

\textbf{Long-term} & $>28$ &
Strategic resource and roster planning &
\begin{itemize}[leftmargin=*, nosep, topsep=0pt]
  \item \textit{As-of} booking counts at coarse thresholds ($\tau \geq 8$ weeks)
\end{itemize}
\\

\bottomrule
\end{tabularx}
\end{table}

\subsection{Evaluation metrics}

Forecast performance is evaluated using a small set of complementary metrics designed to reflect both statistical accuracy and operational risk.
In the context of passenger assistance planning, under-prediction of demand carries a higher operational cost than over-prediction, as insufficient staffing increases the risk of failed assistance delivery, whereas over-staffing primarily results in inefficiency and marginal cost.
The selected metrics explicitly account for this asymmetry.

\paragraph{Mean Absolute Error (MAE).}
Mean Absolute Error provides a scale-dependent measure of average forecast accuracy and is defined as
\begin{equation}
\mathrm{MAE} = \frac{1}{N} \sum_{t=1}^{N} \left| y_t - \hat{y}_t \right|,
\end{equation}
where $y_t$ denotes the observed number of passenger assistance events and $\hat{y}_t$ the corresponding forecast.
MAE is robust to outliers and provides an interpretable measure of typical absolute deviation between forecasts and observed demand.

\paragraph{Asymmetric Root Mean Squared Error (aRMSE).}
To reflect the higher operational risk associated with under-prediction, we employ an asymmetric variant of the Root Mean Squared Error in which negative forecast errors are penalised more heavily than positive errors.
Let $e_t = \hat{y}_t - y_t$ denote the forecast error at time $t$.
The asymmetric RMSE is defined as
\begin{equation}
\mathrm{aRMSE} = \sqrt{
\frac{1}{N} \sum_{t=1}^{N}
w(e_t)\, e_t^2
},
\end{equation}
where the weighting function $w(e_t)$ is given by
\begin{equation}
w(e_t) =
\begin{cases}
2, & \text{if } e_t < 0 \quad \text{(under-prediction)}, \\
1, & \text{if } e_t \geq 0 \quad \text{(over-prediction)}.
\end{cases}
\end{equation}
This formulation explicitly prioritises the reduction of under-forecasting errors, aligning the evaluation metric with the asymmetric costs faced in operational workforce planning.

\paragraph{Coverage probability within tolerance.}
In addition to point-error metrics, we evaluate forecast adequacy using a coverage-based measure that reflects whether forecasts fall within an acceptable tolerance band around observed demand.
For each station $s$, a station-specific tolerance threshold $\delta_s$ is defined to reflect local operational capacity and staffing flexibility.
Coverage probability is then computed as
\begin{equation}
\mathrm{Coverage}_s =
\frac{1}{N_s}
\sum_{t=1}^{N_s}
\mathbb{I}
\left(
\left| y_{s,t} - \hat{y}_{s,t} \right| \leq \delta_s
\right),
\end{equation}
where $\mathbb{I}(\cdot)$ denotes the indicator function and $N_s$ is the number of evaluated time periods for station $s$.
This metric captures the proportion of forecasts that are sufficiently accurate for operational planning purposes, recognising that acceptable error margins vary across stations with different demand profiles.

Together, these metrics provide a balanced assessment of forecast performance, combining average accuracy, asymmetric risk sensitivity, and operational adequacy.

\section{From forecasts to workforce planning}
\label{sec:planning}

\subsection{Planning objectives and constraints}

The objective of the workforce planning component is to translate hourly passenger assistance demand forecasts into an assessment of whether the base station roster provides sufficient capacity to meet expected demand.
The approach is designed as a decision-support tool rather than an optimisation model, with the aim of identifying future periods where staffing levels may be under pressure and where proactive intervention may be required.

Passenger assistance delivery at stations is constrained by the structure of the base roster and by the differing responsibilities of station staff.
Some roles are primarily dedicated to passenger assistance, while others may support assistance delivery only when operationally available and not engaged in competing duties.
Staff availability may also vary by hour within a shift, reflecting dispatch responsibilities, customer service activity, or other operational demands.
In addition, workforce planning must allow for a degree of operational resilience to account for uncertainty, training requirements, and unplanned events.

Given these constraints, the planning logic focuses on estimating a safe and effective hourly assistance capacity for each station and comparing this capacity against forecast peak demand on an hour-by-hour basis.

\subsection{Mapping hourly demand forecasts to staffing capacity}

Forecasts are generated at an hourly resolution, producing an estimate of expected passenger assistance demand for each station-hour.
For workforce planning purposes, attention is focused on the busiest forecast hours, as staffing adequacy is driven by peak load rather than average demand.

Hourly staffing capacity is derived directly from the base roster by identifying which staff roles are scheduled to be working during each hour.
Staff are grouped into two categories:
(i) \emph{primary} staff whose core responsibilities include passenger assistance, and
(ii) \emph{secondary} staff who can provide assistance support subject to availability.

A capacity factor is defined to reflect safe working limits and operational resilience:
\begin{equation}
CF = A_h \times (1 - M),
\end{equation}
where $A_h$ denotes the maximum safe number of assistance events that a fully available staff member can deliver in one hour, and $M$ is an operational margin accounting for uncertainty, training levels, and disruption risk.

Primary hourly capacity is calculated as
\begin{equation}
C_{P,h} = N_{P,h} \times CF,
\end{equation}
where $N_{P,h}$ is the number of primary staff rostered and available during hour $h$.

Secondary hourly capacity is calculated by discounting staff availability:
\begin{equation}
C_{S,h} = \sum_i N_{i,h} \times \alpha_i \times CF,
\end{equation}
where $N_{i,h}$ denotes the number of staff in secondary role $i$ rostered during hour $h$, and $\alpha_i \in [0,1]$ represents the proportion of time those staff are realistically available to support passenger assistance.

Total effective hourly capacity is then given by
\begin{equation}
C_{\text{total},h} = C_{P,h} + C_{S,h}.
\end{equation}

\subsection{Hourly workforce planning logic and alert classification}

For each station-hour, forecast demand is compared directly against calculated staffing capacity using a simple red--amber--green (RAG) alert framework.
Let $\hat{y}_{s,h}$ denote the forecast assistance demand for station $s$ during hour $h$.

Hourly alert levels are defined as follows:
\begin{itemize}
    \item \textbf{Green:} $\hat{y}_{s,h} \leq C_{P,h}$, indicating that forecast demand can be met using primary assistance staff alone.
    \item \textbf{Amber:} $C_{P,h} < \hat{y}_{s,h} \leq C_{\text{total},h}$, indicating reliance on secondary staff and elevated operational pressure.
    \item \textbf{Red:} $\hat{y}_{s,h} > C_{\text{total},h}$, indicating that forecast demand exceeds available staffing capacity and that intervention is likely to be required.
\end{itemize}

This hourly classification provides a transparent and intuitive assessment of staffing risk that aligns directly with how station rosters are constructed and managed.
Amber and red alerts highlight specific hours where proactive actions may be required, such as short-term redeployment, targeted overtime, or local operational adjustments. By operating at an hourly level, the framework captures intraday variation in both demand and staffing availability, allowing planners to distinguish between brief peak pressures and sustained resourcing shortfalls.
This supports more targeted and proportionate workforce interventions while maintaining a clear link between forecast demand and operational decision-making.

\section{Results}
\label{sec:results}
\subsection{Representative stations}
\label{sec: rep stations}

To illustrate model performance and operational implications across different demand regimes, results are presented for three representative stations selected from the thirteen stations that are managed by LNER:

\begin{itemize}
    \item \textbf{London King's Cross} represents a high-volume long-distance origin and destination station, characterised by strong weekly and seasonal patterns and pronounced peak demand during holiday periods.

    \item \textbf{York} represents a major interchange hub, where a substantial share of passenger assistance demand arises from transferring passengers and demand is more tightly clustered around connection times.

    \item \textbf{Berwick-upon-Tweed} represents a low-volume, local station with sparse and intermittent assistance demand, including extended periods with zero recorded events.
\end{itemize}

These stations capture the principal operational contexts observed across the wider station portfolio and illustrate performance under contrasting demand conditions.

\subsection{Forecasting performance}
\label{sec:results_performance}
For each horizon bucket, Prophet models were trained using a time-ordered training validation framework to preserve the temporal structure of the data. Hyperparameters governing trend flexibility, seasonal regularisation, and holiday effects were tuned using a grid search. In particular, seasonal flexibility was controlled via the seasonality prior scale and holidays prior scale, both tuned over the range 0.01–10 to regulate the magnitude of recurring effects while retaining sufficient responsiveness to peak periods. In addition, both additive and multiplicative seasonality modes were evaluated, with the final choice made based on out-of-sample performance and inspection of residual structure. 

\begin{figure}[!htbp]
\centering
\includegraphics[width=\linewidth]{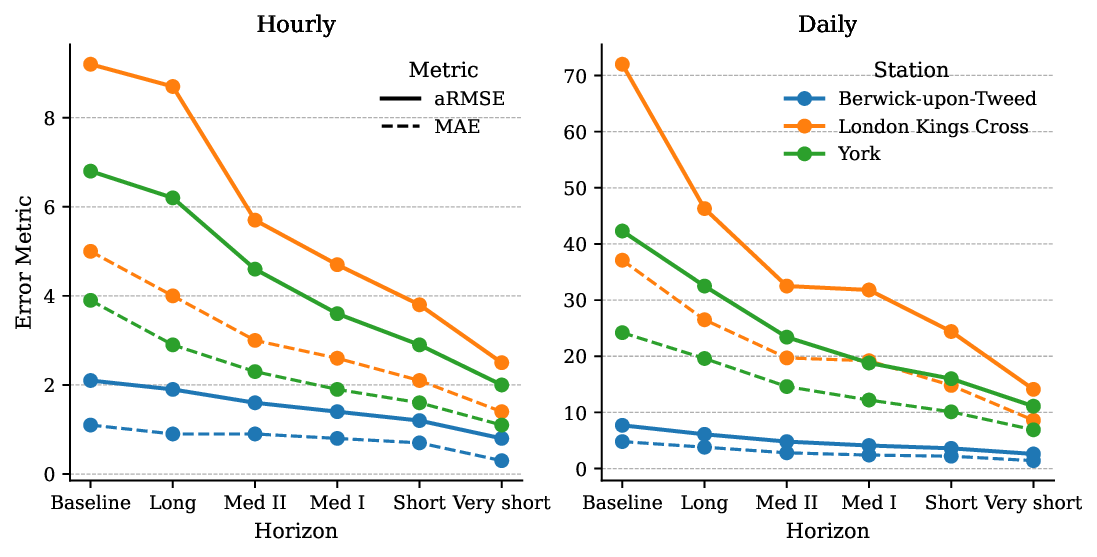}
\caption{Forecast error across horizon buckets for representative stations at hourly (left) and daily (right) resolutions. Solid lines denote asymmetric RMSE and dashed lines denote MAE.}
\label{fig:error_by_horizon}
\end{figure}

Figure~\ref{fig:error_by_horizon} summarises forecast accuracy across horizon buckets for the three representative stations, shown separately for hourly (left panel) and daily (right panel) resolutions. Solid lines denote asymmetric root mean squared error (aRMSE), while dashed lines denote mean absolute error (MAE). 

Across all stations and both temporal resolutions, forecast error decreases monotonically as the prediction horizon shortens. This reflects the increasing availability of booking information closer to the date of travel, which strengthens the predictive signal and reduces uncertainty. Quantitatively, very short-term forecasts reduce aRMSE relative to the year-on-year baseline by approximately $66\%$ at London King’s Cross, $74\%$ at York, and $66\%$ at Berwick-upon-Tweed at the daily level, with comparable proportional improvements observed at the hourly resolution. Similar reductions are observed for MAE, confirming that improvements are not driven solely by the asymmetric penalty structure of the evaluation metric.

From an operational perspective, this pattern is important. As the forecast horizon shortens, staffing flexibility becomes more limited and the cost of underestimation increases, as late adjustments tend to rely on reactive and often more expensive interventions. Improvements in near-term forecast accuracy are therefore particularly valuable, as they provide more reliable demand visibility at the point where day-to-day staffing decisions are made and the risk of service failure is highest.

Differences in absolute error magnitudes reflect the heterogeneous demand regimes outlined in Section \ref{sec: rep stations} and demonstrate that the framework performs consistently across contrasting operational contexts. As expected, London King’s Cross exhibits the largest absolute errors due to its higher volumes, but also the greatest absolute gains as the horizon shortens. York shows steady improvements across horizons, while Berwick-upon-Tweed maintains low error levels with a clear monotonic pattern despite its sparse and intermittent demand. Taken together, these results indicate that the modelling approach remains stable across high-volume, hub, and low-volume stations, suggesting that performance improvements are not driven by a single demand profile but generalise across the station portfolio.

% --- DAILY TABLE ---
\begin{table}[!htbp]
\centering
\caption{Daily forecast performance by station and horizon (in days).}
\label{tab:daily_performance_full}
\begin{tabular}{llccc}
\toprule
Station & Horizon (days) & MAE & aRMSE & Coverage (\%) \\
\midrule
\multicolumn{5}{l}{\textit{Berwick-upon-Tweed}} \\
Baseline (YoY) & Baseline & 4.8 & 7.7 & 100.0 \\
Long Term & $>28$ & 3.8 & 6.1 & 100.0 \\
Medium II Term & 15--28 & 2.8 & 4.8 & 100.0 \\
Medium I Term & 8--14 & 2.4 & 4.1 & 100.0 \\
Short Term & 3--7 & 2.2 & 3.6 & 100.0 \\
Very Short Term & 1--2 & 1.4 & 2.6 & 100.0 \\
\addlinespace
\multicolumn{5}{l}{\textit{London King’s Cross}} \\
Baseline (YoY) & Baseline & 37.1 & 72.0 & 58.4 \\
Long Term & $>28$ & 26.5 & 46.3 & 71.4 \\
Medium II Term & 15--28 & 19.7 & 32.5 & 77.8 \\
Medium I Term & 8--14 & 19.2 & 31.8 & 82.2 \\
Short Term & 3--7 & 14.8 & 24.4 & 89.9 \\
Very Short Term & 1--2 & 8.6 & 14.1 & 96.3 \\
\addlinespace
\multicolumn{5}{l}{\textit{York}} \\
Baseline (YoY) & Baseline & 24.2 & 42.3 & 72.0 \\
Long Term & $>28$ & 19.6 & 32.5 & 78.8 \\
Medium II Term & 15--28 & 14.6 & 23.4 & 87.2 \\
Medium I Term & 8--14 & 12.2 & 18.8 & 92.9 \\
Short Term & 3--7 & 10.1 & 16.0 & 94.9 \\
Very Short Term & 1--2 & 6.9 & 11.1 & 99.7 \\
\bottomrule
\end{tabular}
\end{table}

% --- HOURLY TABLE ---
\begin{table}[!htbp]
\centering
\caption{Hourly forecast performance by station and horizon (in days).}
\label{tab:hourly_performance_full}
\begin{tabular}{llccc}
\toprule
Station & Horizon (days) & MAE & aRMSE & Coverage (\%) \\
\midrule
\multicolumn{5}{l}{\textit{Berwick-upon-Tweed}} \\
Baseline (YoY) & Baseline & 1.1 & 2.1 & 98.1 \\
Long Term & $>28$ & 0.9 & 1.9 & 99.2 \\
Medium II Term & 15--28 & 0.9 & 1.6 & 99.9 \\
Medium I Term & 8--14 & 0.8 & 1.4 & 100.0 \\
Short Term & 3--7 & 0.7 & 1.2 & 100.0 \\
Very Short Term & 1--2 & 0.3 & 0.8 & 100.0 \\
\addlinespace
\multicolumn{5}{l}{\textit{London King’s Cross}} \\
Baseline (YoY) & Baseline & 5.0 & 9.2 & 65.4 \\
Long Term & $>28$ & 4.0 & 8.7 & 74.6 \\
Medium II Term & 15--28 & 3.0 & 5.7 & 81.1 \\
Medium I Term & 8--14 & 2.6 & 4.7 & 86.1 \\
Short Term & 3--7 & 2.1 & 3.8 & 91.0 \\
Very Short Term & 1--2 & 1.4 & 2.5 & 97.9 \\
\addlinespace
\multicolumn{5}{l}{\textit{York}} \\
Baseline (YoY) & Baseline & 3.9 & 6.8 & 72.4 \\
Long Term & $>28$ & 2.9 & 6.2 & 84.0 \\
Medium II Term & 15--28 & 2.3 & 4.6 & 89.2 \\
Medium I Term & 8--14 & 1.9 & 3.6 & 93.2 \\
Short Term & 3--7 & 1.6 & 2.9 & 96.3 \\
Very Short Term & 1--2 & 1.1 & 2.0 & 99.0 \\
\bottomrule
\end{tabular}
\end{table}

Tables~\ref{tab:daily_performance_full} and~\ref{tab:hourly_performance_full} report performance metrics across the full set of horizon buckets for daily and hourly forecasts. Presenting results in this way reflects the operational setting in which forecasts are used at multiple decision points rather than evaluated at a single lead time. When interpreted alongside the planning responsibilities described in Section \ref{sec: workforce planning and constraints} and summarised in Table \ref{tab:planning_horizons}, the progression of forecast accuracy aligns closely with how decisions are made in practice. Long-horizon forecasts, used for strategic and resource planning, exhibit higher uncertainty but still deliver meaningful improvements over baseline performance, providing directional guidance where planning flexibility is greatest. Medium-horizon forecasts, corresponding to roster clerks and accessibility planning teams, show substantial reductions in both MAE and aRMSE, supporting tactical adjustments while roster modifications remain feasible. Short and very short-term forecasts, used by service delivery teams and station managers, achieve the lowest errors and highest coverage levels, providing the level of reliability required for day-of-operation staffing decisions where the operational cost of underestimation is highest. 

\subsection{Workforce planning RAG heatmap}
\label{subsec:workforce_rag_results}

\begin{figure}[t]
    \centering
    % Replace with your actual filename/path
    \includegraphics[width=\linewidth]{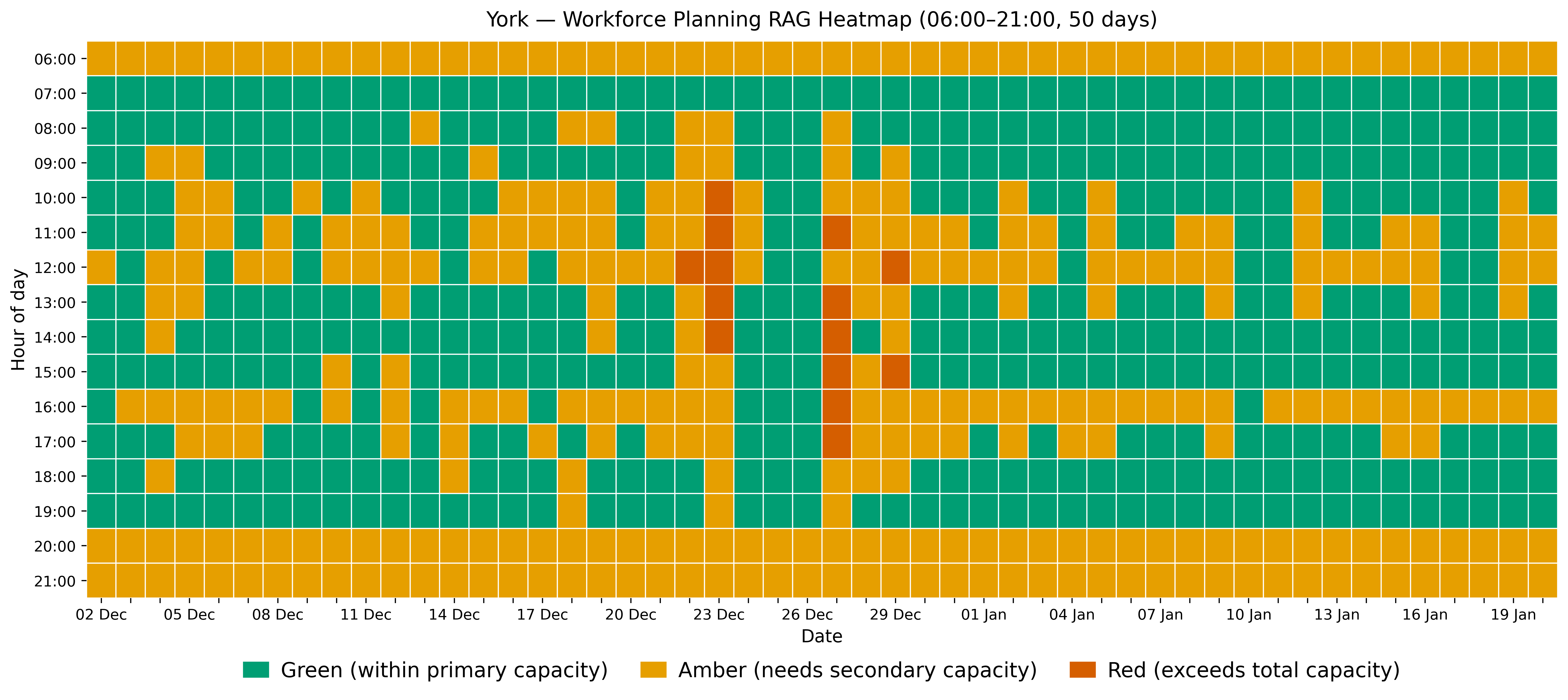}
    \caption{Workforce planning RAG heatmap for York (06:00--21:00, 50 days). Green indicates forecast demand is within primary capacity; Amber indicates secondary capacity is required; Red indicates forecast demand exceeds total capacity.}
    \label{fig:rag_heatmap_york}
\end{figure}

Following the methodology introduced in Section~\ref{sec:planning}, we combine the demand forecasts with an hourly baseline workforce plan to visualise operational risk over time. Figure~\ref{fig:rag_heatmap_york} presents the resulting Red--Amber--Green (RAG) heatmap for York, where each cell represents a station-hour and is classified as \textit{Green} (forecast demand within primary capacity), \textit{Amber} (demand can be met only by drawing on secondary capacity), or \textit{Red} (demand exceeds total available capacity).

The base roster is specified at an hourly level and decomposed into primary and secondary roles. Performance Station Assistants (PSA) and Station Support Customer Service (SCSC) staff are treated as primary because they are principally responsible for delivering Passenger Assist. Station Customer Service Assistants (SCSA), whose primary responsibility is dispatch, and Station Support Assistants (SSA), who operate help-point kiosks, are treated as secondary because providing assist support requires them to step away from their core duties. Roles that may only assist as a last resort, such as Information Controllers and Duty Team Leaders, are excluded from the capacity calculation to ensure that estimated capacity reflects routinely available operational resource.

Capacity is parameterised using a maximum handling rate of $A_h=4$ assists per hour per fully available staff member, with an operational margin of $M=0.10$ to preserve resilience against stochastic variation and unmodelled disruptions. Secondary staff contribute at a reduced effectiveness of $\alpha=0.30$. This yields an effective capacity of $A_h(1-M)=3.6$ assists per hour for primary staff and $1.08$ assists per hour for secondary staff when their support is required.
The heatmap highlights clear periods of elevated risk around the Christmas travel peak. The demand surge on 23\textsuperscript{rd} December produces several \textit{Red} hours, indicating that even after incorporating discounted secondary capacity, forecast demand exceeds total available capacity. A similar pattern is observed on 27\textsuperscript{th} December, where multiple hours shift into \textit{Amber} and \textit{Red}. Importantly, the approach also identifies 29\textsuperscript{th} December as a substantial risk period despite being outside the most obvious holiday peak, demonstrating the value of the framework in providing forward visibility of non-trivial high-demand days. Beyond the holiday period, the RAG visualisation reveals recurring baseline pressure in the existing roster. In the first weeks of the New Year, the 10:00--13:00 window consistently appears as \textit{Amber} on Mondays, indicating a systematic reliance on secondary staff even under non-exceptional demand conditions.

\section{Discussion}
\label{sec:discussion}
\subsection{Explainability}

A key advantage of the Prophet modelling framework is its inherent
interpretability. As described in Section~\ref{sec:prophet}, the forecasting approach
is formulated as an additive, Generalised Additive Model (GAM)-style
decomposition in which demand is represented as the sum of trend,
seasonal, holiday, and exogenous regressor components. This structural
formulation provides a direct link between the statistical model and the
operational drivers of demand. Unlike many black-box approaches, each
prediction can therefore be traced back to its underlying components,
providing transparency that is particularly valuable in a
decision-support setting.

\begin{figure}[htbp]
    \centering
    \includegraphics[width=\textwidth]{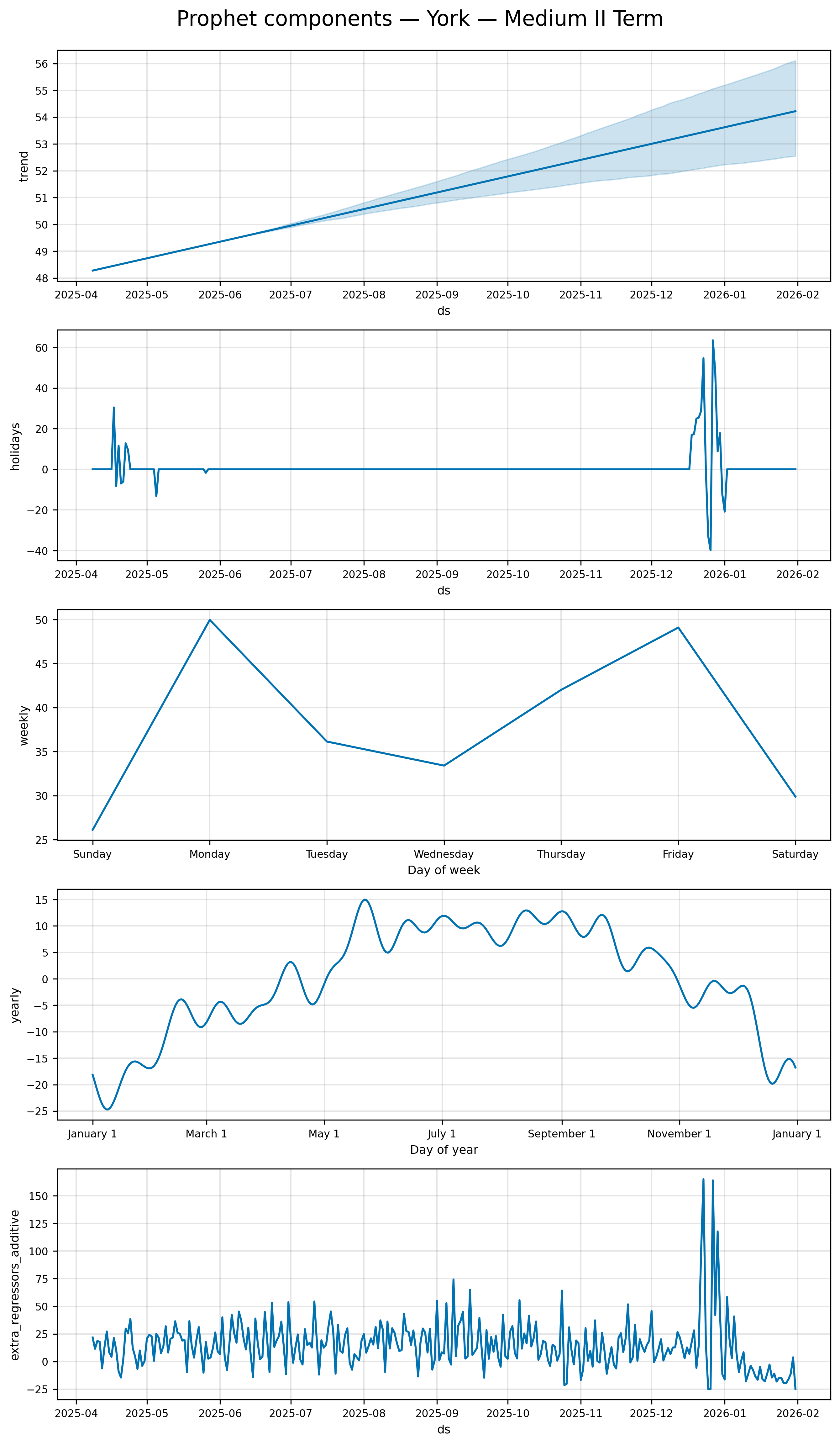}
    \caption{Prophet component decomposition for York (Medium II horizon),
    illustrating the contribution of trend, multiple seasonalities,
    holiday effects, and external regressors to the final forecast.}
    \label{fig:prophet_components_york}
\end{figure}

Figure~\ref{fig:prophet_components_york} illustrates how a representative
forecast for York (Medium II horizon) can be decomposed into its
constituent elements. The long-term trend component captures gradual
structural changes in assistance demand, reflecting factors such as
growth in service uptake or underlying passenger volumes. Superimposed
on this are seasonal effects at multiple temporal scales. The weekly
component highlights strong day-of-week patterns consistent with
commuting and leisure travel behaviours, while the yearly component
reflects broader seasonal cycles such as summer travel peaks and reduced
post-holiday demand. The intraday (daily) component captures pronounced
within-day structure, with demand concentrated around typical travel and
interchange periods.

Holiday effects provide an additional interpretable adjustment,
enabling the model to represent demand perturbations during public
holidays and major travel periods. Importantly, the inclusion of external
regressors (most notably cumulative booking features) allows forecasts to
respond dynamically to real-time demand signals. The regressor
contribution panel shows how spikes in bookings directly translate into
forecast adjustments. Together, these components provide a transparent
explanation of how the model arrives at a given prediction, enabling
planners to understand whether a forecast is driven primarily by
structural seasonality, calendar effects, or emerging booking activity.

\begin{figure}[htbp]
    \centering
    \includegraphics[width=\textwidth]{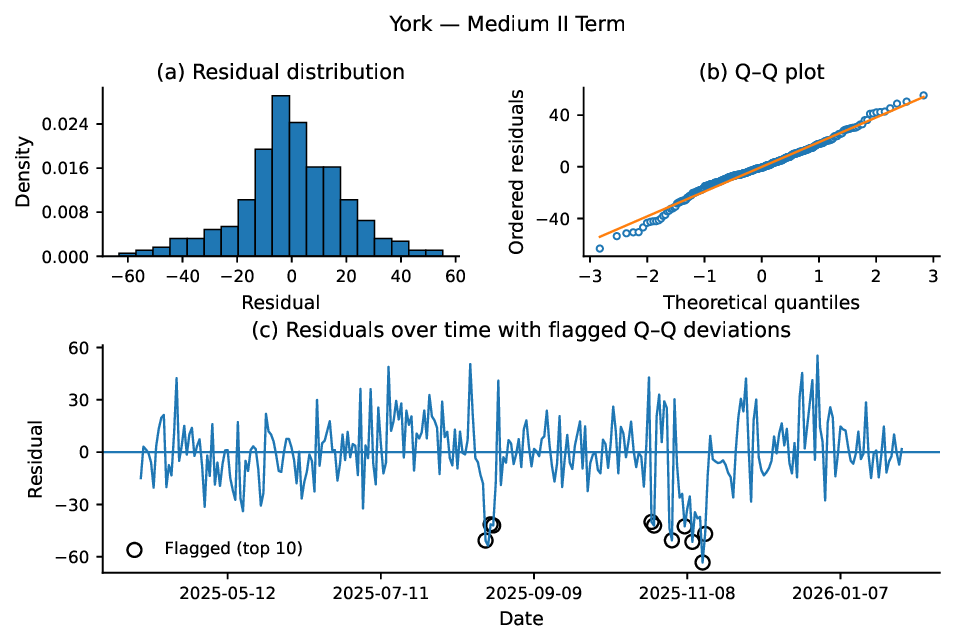}
    \caption{Residual diagnostics for York (Medium II horizon), including
    the residual distribution, Q--Q plot, and residuals over time.
    Circled observations correspond to periods of engineering works not
    explicitly captured in the model.}
    \label{fig:residual_analysis_york}
\end{figure}

Figure~\ref{fig:residual_analysis_york} presents diagnostic plots of
forecast errors, including the residual distribution, Q--Q plot, and the
temporal evolution of residuals. The residual histogram and Q--Q plot
indicate that errors are broadly centred and approximately symmetric,
suggesting that the model captures the dominant demand structure.
However, the time-series residual plot reveals several pronounced
deviations, highlighted by the circled observations.

Operational review of these flagged periods showed that they coincide
with major engineering works affecting services through York leading to a limited service for some operators. Such
events disrupt typical travel patterns and assistance demand but are not
explicitly represented in the current model specification. The presence
of these systematic deviations therefore provides actionable insight:
rather than indicating random model failure, they identify specific
exogenous shocks that could be incorporated as future regressors or
event indicators.

This combination of component-level interpretability and residual
diagnostics strengthens confidence in the forecasting framework.
Planners can not only see the expected demand but also understand why the
model predicts it and when deviations are likely to arise due to
unmodelled operational events. In practice, this transparency supports
stakeholder trust and provides a clear pathway for iterative model
improvement, such as the inclusion of disruption or engineering-work
indicators in future versions of the system.
\subsection{Impact Since Production}
The forecasting and decision-support framework was deployed into the operational planning workflow in December 2025. Although only a limited post-deployment period is currently available for evaluation, early operational evidence indicates meaningful improvements in both workforce planning effectiveness and service reliability. The mechanism is direct: improved forecast accuracy provides earlier visibility of periods in which assistance demand is likely to exceed the baseline roster, enabling planners and station teams to take mitigating actions such as targeted overtime, staff redeployment, or short-term roster adjustments. In the initial months following release, use of the system was associated with an approximate 50\% reduction in failed passenger assistance deliveries attributable to staff availability (measured per 1{,}000 assists) compared with the equivalent pre-deployment period.

A particularly illustrative example occurred during the post-Christmas
operating period. Historically, the first operating day following
Christmas (typically 27th December) experiences a pronounced spike in
assistance demand. In the first year of deployment, however, the 27th
fell on a Saturday, leading to a redistribution of demand across the
27th–29th December rather than a single-day peak. The YoY growth
forecasting approach did not capture this calendar-driven shift and
would have underestimated demand on 29th December by more than 700
assistance bookings across all thirteen LNER stations. In contrast, the deployed forecasting framework identified the emerging demand pattern and reduced forecast error by
approximately 93\% relative to the static baseline. This improved
visibility enabled planners to adjust staffing and operational
preparations accordingly, mitigating the risk of service shortfalls
during a period characterised by high passenger volumes and a large
proportion of infrequent travellers. While these results represent early evidence rather than a long-term
evaluation, they demonstrate the practical value of embedding
data-driven forecasting into routine operational decision processes.

\section{Conclusions and future work}
\label{sec:conclusion}
This paper presented a data-driven framework for forecasting station-level passenger assistance demand and translating these forecasts into actionable workforce planning insights. By combining a decomposable time-series modelling approach with horizon-specific regressors, the framework captures both the strong temporal structure of assistance demand and the evolving signal provided by booking behaviour and external conditions. The horizon-bucket methodology enables forecasts to remain stable and interpretable at longer lead times while exploiting richer information as the date of operation approaches, aligning closely with real-world planning horizons.

Empirical evaluation demonstrates consistent improvements in forecast accuracy across stations with markedly different demand profiles, including high-volume terminals, interchange hubs, and low-volume local stations. Error reductions increase as the prediction horizon shortens, reflecting the growing predictive value of near-term booking information. When integrated with the workforce planning component, the forecasts provide clear visibility of capacity risk through an intuitive hourly RAG framework, enabling planners to identify peak pressure periods and take proactive mitigation actions. Early operational evidence following deployment further indicates meaningful improvements in service reliability and planning effectiveness, highlighting the practical value of embedding predictive analytics within routine operational workflows.

While the framework captures the principal drivers of assistance demand, several extensions offer opportunities for further improvement. A key direction is the incorporation of additional exogenous data sources that reflect atypical operating conditions. In particular, structured information on planned engineering works could be integrated as event-based regressors, enabling the model to account explicitly for temporary service alterations that can materially affect demand patterns as depicted in Figure~\ref{fig:residual_analysis_york}. Similarly, incorporating data on large-scale events such as concerts, conferences, and major sporting fixtures would allow the forecasting system to anticipate location-specific demand surges more effectively, particularly at stations serving event venues or key interchange routes. The inclusion of real-time and historical disruption indicators, such as delays, cancellations, or broader network performance measures, represents another promising avenue, with potential to improve short-term responsiveness and support scenario-based planning during periods of operational instability.

Certain parameters within the staffing model, such as maximum assists per staff member per hour and resilience margins, are based on operational expertise. While these provide a practical and interpretable framework for decision support, future work could explore calibration and sensitivity analysis to further validate and refine these assumptions.

While this study focuses on LNER-managed stations, the approach is transferable across operators. Given the interconnected nature of Passenger Assist journeys, wider adoption could support more coordinated accessibility planning across the UK rail network, improving consistency and reliability of assisted travel at a system level

\section*{Acknowledgements}
The author would like to thank colleagues within London North Eastern Railway for their support in the development and deployment of this work. In particular, the contribution of the Accessibility team and the Data Services team is gratefully acknowledged for their domain expertise, insight, and ongoing collaboration.
% ---------------- References ----------------
\bibliographystyle{elsarticle-harv}
\bibliography{references}

% ---------------- Appendices ----------------
\appendix

\end{document}